\documentclass[prb,twocolumn,showpacs,superscriptaddress,preprintnumbers,amsmath,amssymb]{revtex4}

\usepackage{graphicx}
\usepackage{amssymb,amsmath}
\usepackage{dcolumn}
\usepackage{bm}
\usepackage{color}
\usepackage{enumerate}

\begin{document}

\title{THz laser based on dipolaritons}

\author{K. Kristinsson}
\affiliation{Division of Physics and Applied Physics, Nanyang Technological University 637371, Singapore}

\author{O. Kyriienko}
\affiliation{Division of Physics and Applied Physics, Nanyang Technological University 637371, Singapore}
\affiliation{Science Institute, University of Iceland, Dunhagi-3, IS-107, Reykjavik, Iceland}

\author{I. A. Shelykh}
\affiliation{Division of Physics and Applied Physics, Nanyang Technological University 637371, Singapore}
\affiliation{Science Institute, University of Iceland, Dunhagi-3, IS-107, Reykjavik, Iceland}

\date{\today}

\begin{abstract}
We develop the microscopic theory of a terahertz (THz) laser based on the effects of resonant tunneling in a double quantum well heterostructure embedded in both optical and THz cavities. In the strong coupling regime the system hosts dipolaritons, hybrid quasiparticles formed by the direct exciton, indirect exciton and optical photon, which possess large dipole moments in the growth direction. Their radiative coupling to the mode of a THz cavity combined with strong non-linearities provided by exciton-exciton interactions allows for stable emission of THz radiation in the regime of the continuous optical excitation. The optimal parameters for maximizing the THz signal output power are analyzed.
\end{abstract}

\maketitle

\section{INTRODUCTION}

The possibility of experimental design of systems of reduced
dimensionality has allowed the study of novel types of
quasi-particles absent in bulk structures. One of the examples is an
indirect exciton, a bound state formed by an electron and a hole
confined in spatially separated quantum wells
(QWs).\cite{Lozovik,ButovReview} Its energy can be effectively tuned
by an electric field applied perpendicular to the structure's
interface.\cite{Butov2001} Due to the small overlap between the
wavefunctions of an electron and a hole, indirect excitons have a
much longer radiative lifetime compared to direct excitons, for
which an electron and a hole are localized in the same
QW.\cite{Butov2001,Alexandrou} Moreover, the presence of an inherent
dipole moment strongly enhances exciton-exciton
interactions,\cite{Rapaport, Schindler} making the system of
indirect excitons an attractive candidate for the experimental
observation of quantum collective phenomena, including excitonic
Bose-Einstein condensation (BEC).\cite{High2012,Butov2002,Snoke2002}

Another microstructure in which formation of macroscopically
coherent states has been reported experimentally is a semiconductor
microcavity in the strong coupling regime,
\cite{Microcavities,CarusottoRev} in which the optical cavity mode
effectively hybridizes with the excitonic mode of a QW embedded in
the antinode position. The elementary excitations in this case are
cavity polaritons, bosonic quasiparticles containing both light and
material fractions. Due to their extremely low effective mass
inherited from the cavity photon part a non-equilibrium BEC of
polaritons was observed in high quality cavities at liquid
nitrogen\cite{Kasprzak,Balili,SvenNature} and even at room
temperature.\cite{Christopoulos2007} Other collective phenomena
observed in microcavities include superfluidity,\cite{Amo2009}
Josephson effect,\cite{Lagoudakis} optical
bistability,\cite{Baas2004} condensate
phase-locking,\cite{Cristofolini2013} formation of quantized
vortices, \cite{Lagoudakis2011} and
solitons.\cite{Sich2011,Hivet2012,Egorov2010}

\begin{figure}[t!]
\includegraphics[width=0.44\textwidth]{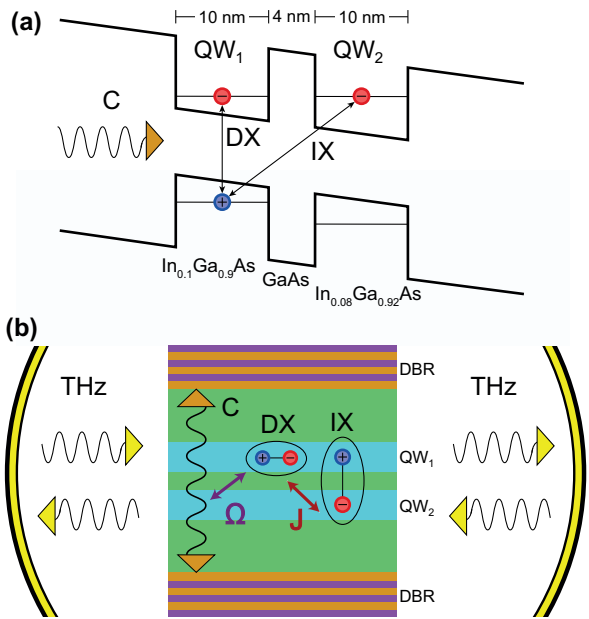}
\caption{(Color online) (a) Band diagram of the double QW system. In the presence of the optical microcavity the QW$_1$ is coupled to the cavity mode. The energy bands are tilted by the applied electric field, bringing the electron levels into resonance. Dimensions and materials of the QWs and the barrier are indicated. (b) Sketch of the dipolariton system in a THz cavity. The QW$_1$ is where direct excitons (DX) are excited. The electron can tunnel with a rate $J$ to the QW$_2$, forming indirect excitons (IX). The distributed Bragg reflectors (DBR) provide the confinement of the cavity photon (C), which is strongly coupled to the direct exciton transition with a Rabi frequency $\Omega$, and decoupled from the QW$_2$. The supplemental cavity hosts a THz photonic mode, which interacts with the exciton dipole moment. The bare double QW system lacks the DBR optical confinement, and therefore the microcavity mode.}
\label{fig:sketch}
\end{figure}

Indirect excitons and polaritons can be combined in a hybrid structure consisting of a pair of non-equivalent QWs inside a planar microcavity.\cite{Cristofolini2012,Christmann2011,Christmann2010,Muljarov2012} If the bandgaps of the two QWs are different, the cavity mode can be tuned to couple resonantly to the excitonic transition in only one of them. In the same time, tuning electron levels of the two QWs into resonance by applying the external electric field, one can achieve the strong tunneling coupling between a direct exciton and a spatially indirect exciton formed by an electron and a hole located in different QWs [Fig. \ref{fig:sketch}(a)]. These strong couplings lead to the appearance of new eigenmodes of the system, which represent three linear superpositions of the cavity photon (C), direct exciton (DX) and indirect exciton (IX) modes. They are called the upper dipolariton (UP), the middle dipolariton (MP) and the lower dipolariton (LP).

It was recently proposed by the authors of the present paper that dipolariton system can be used as a source of the THz emission in the regimes of both pulsed \cite{Kyriienko2013} and continuous \cite{Kristinsson} excitation. The effect was based on the possibility of achieving huge alternations of the dipole moment in time domain arising from Rabi oscillations between direct and indirect excitons.  The important drawback of the approach used in Refs. [\onlinecite{Kyriienko2013,Kristinsson}] was that the THz emission was considered in purely classical manner, which excluded the possibility of the analysis of the phenomena related to the possible onset of THz lasing.

In the present paper we account for the quantum nature of the THz emission and propose an experimentally-friendly compact scheme of a dipolariton-based laser, operating in a THz domain of frequencies, shown in Fig. \ref{fig:sketch}(b). To enhance the positive feedback for a THz emission we place the double well system inside a high-quality THz cavity supporting a mode polarized perpendicular to the QW planes to maximize its coupling to the excitonic dipoles.

The paper is organized as follows. We first introduce the
Hamiltonian of the dipolariton system, accounting for the THz mode
and its interaction with the excitonic modes, and derive
corresponding equations of motion for the fields. We then
demonstrate the difference between the bare double QW system, in
which oscillations are strongly anharmonic, and the full dipolariton
system, where oscillations are harmonic, and show that harmonicity
is vital for high stationary occupation of the THz mode. Next we
calculate the emission power and investigate its dependence on the
THz cavity quality factor $Q$, finding Purcell enhancement at low
$Q$, and weak dependence past some critical value. Furthermore, we
demonstrate coherence-induced superradiance of the emission, which
strongly improves the efficiency of the emitter, and address the
issue of tunability of the emission frequency.

\section{DOUBLE QUANTUM WELL SYSTEM IN THZ CAVITY}
In the first part of the paper we consider the bare double QW system
placed inside a THz cavity, assuming an optical cavity being absent
[as in Fig. \ref{fig:sketch}(b), but without DBRs]. The upper
quantum well (QW$_1$) hosts direct excitons (DX) with bare energy
denoted by $\hbar \omega_D$. The electrons can tunnel  to the lower
quantum well (QW$_2$) with a rate $J$, forming indirect excitons
(IX) with energy $\hbar \omega_I$. Note that the detuning between
direct and indirect excitons $\delta_J =\omega_I -\omega_D$ can be
tuned by the applied electric field $F$.

We now put the double QW system inside the THz optical cavity of the
wavelength $\lambda_T$, with the corresponding frequency $\omega_T$.
Using the second quantization formalism we introduce the bosonic
creation and annihilation operators for the direct ($\hat b
^\dagger$, $\hat b$) and indirect excitons ($\hat c^\dagger$, $\hat
c$), as well as the THz cavity photons ($\hat a^{\dagger}$, $\hat
a$). The Hamiltonian of the system can be written as a sum of free
energy and tunneling terms, exciton-THz mode interaction, nonlinear
Coulomb interaction and pumping terms,
\begin{align}
\hat{\mathcal H} = \hat{\mathcal H}_0 +\hat{\mathcal H}_{int} +\hat{\mathcal H}_{nonl} +P(t){\hat b}^\dagger+P(t)^* {\hat b},
\label{full_no_cav}
\end{align}
where two last terms describe the coherent optical pumping of the DX
mode with intensity $|P(t)|^2$. For the coherent continuous wave
(CW) pumping its time dependence has the form $P(t) = P_0
e^{-i\omega_p t}$, where $\hbar\omega_p$ is the pump energy, and
$P_0$ is the amplitude.

The first term reads
\begin{align}
\hat{\mathcal H}_0 =& \hbar \omega_T {\hat a}^\dagger{\hat a}+\hbar \omega_D {\hat b}^\dagger{\hat b}+\hbar \omega_I {\hat c}^\dagger{\hat c}-\frac{\hbar J} 2 ({\hat b}^\dagger{\hat c}+{\hat c}^\dagger{\hat b}), \label{H0}
\end{align}
and describes the energy of the free mode terms of the THz mode
($\hbar \omega_T$), the direct exciton ($\hbar \omega_D$) and the
indirect exciton ($\hbar \omega_I$), as well as the DX-IX tunneling
with rate $J$.

The third term, corresponding to the nonlinear processes, is given by
\begin{align}
\hat{\mathcal H}_{nonl}= \frac{\alpha_1}{2} \hat b^\dagger \hat b^\dagger \hat b \hat b + \frac{\alpha_3}{2} \hat c^\dagger \hat c^\dagger \hat c \hat c+\alpha_2  \hat b^\dagger \hat c^\dagger \hat b \hat c,\label{Hnonl}
\end{align}
It describes the Coulomb scattering of two direct excitons with
interaction constant $\alpha_1$, two indirect excitons with constant
$\alpha_3$, and the interspecies scattering of a direct and an
indirect exciton  with constant $\alpha_2$. The estimation of these
constants can be found in Ref. [\onlinecite{Kristinsson}].

We can now proceed to account for the interaction between the excitons and the THz cavity, indicated by the second term in Eq. (\ref{full_no_cav}).

\subsection{Interaction Hamiltonian}
In order to include the interaction between DX, IX, and THz modes we
employ the dipole approximation, for which the interaction
Hamiltonian can be represented as
\begin{equation}
\hat{\mathcal H}_{int} = - \hat{\bf d}\cdot \hat{\bf E},
\label{Hint}
\end{equation}
where $\hat{\bf d}$ is the dipole moment operator of the excitonic
system of the double QW, and $\hat{\bf E}$ is the operator of the
electric field corresponding to the THz mode, which in a single mode
approximation can be represented as:
\begin{equation}
\hat{\bf E} = \sqrt{\frac{\hbar \omega_T}{2\epsilon V}} ({\bf e}{\hat a} +{\bf e}^*{\hat a}^\dagger ),
\end{equation}
where $\epsilon$ is the electric permittivity, $V$ is
the cavity volume and
${\bf e}$ is the THz mode polarization vector.

Using the creation and annihilation operators for the direct and
indirect excitons, the dipole moment operator can be written
\begin{align}
\hat{\bf d} = &\:{\bf d}_{dd}\hat b^\dagger \hat b +{\bf d}_{ii} \hat c^\dagger \hat c + {\bf d}_{di}\hat b^\dagger \hat c + {\bf d}_{id}\hat c^\dagger \hat b , \label{dipole_operator}
\end{align}
where ${\bf d}_{jk} = \langle j |\hat{\bf d}|k\rangle$ are the
dipole matrix elements of the double QW exciton states. Because of
the cylindrical symmetry of the system we have ${\bf d}_{jk} =
-e\langle j|\hat{z}|k\rangle {\bf e}_z$, where $e$ denotes the elementary
charge and the $z$-axis is aligned perpendicular to the QW plane.

We can estimate the indirect exciton dipole matrix element as the
electron charge multiplied by an effective electron-hole separation
$L$, $d_{ii}=-e\langle IX|\hat{z}|IX\rangle = -eL$. The direct
exciton dipole element $d_{dd}$ arising from the quantum-confined
Stark effect can be estimated as
\begin{equation}
d_{dd}=-e\langle DX|\hat{z}|DX\rangle = -24 \left( \frac 2 {3\pi}\right)^6 \frac{e^2 F m^* d^4} {\hbar^2},
\end{equation}
where $F$ is the applied electric field, $m^*$ is the quantum well
effective mass and $d$ is the QW width. For a considered
In$_{0.1}$Ga$_{0.9}$As quantum well the effective mass is $m^* =
0.06m_e$, where $m_e$ is the mass of a free electron. The exciton
resonance occurs at $F_0 =12.5$ kV/cm, and the QWs have a thickness
of $d=10$ nm, separated by a $4$ nm barrier. This implies an
approximate electron hole separation of $L=10$ nm. Naturally, the
estimated dipole moment of direct exciton in $z$ direction is much
smaller than that of indirect exciton, $d_{dd}/d_{ii} \approx
10^{-3}$ and is neglected in any further consideration. The
calculation of the $d_{di}$ is presented in the Appendix A, for the
parameters we consider it can be estimated as $d_{di}\approx
0.17d_{ii}$.

Using Eq. (\ref{dipole_operator}) the Hamiltonian (\ref{Hint}) can
be recast as
\begin{align}
\hat{\mathcal H}_{int} =& - [g {\hat c}^\dagger{\hat c}+\tilde g ({\hat b}^\dagger{\hat c}+{\hat c}^\dagger{\hat b})] ({\hat a}+{\hat a}^\dagger), \label{Hint_IX}
\end{align}
where $g= eL \sqrt{\hbar \omega_T / 2\epsilon V}$ and $\tilde g= d_{di} \sqrt{\hbar \omega_T / 2\epsilon V}$ are the exciton-THz photon coupling constants.

To better understand the origin of the THz emission it is useful to
consider the case of zero exciton detuning, $\omega_D = \omega_I$.
In this case the eigenmodes of the bare Hamiltonian $ \hat{\mathcal
H}_0 $ are the symmetric exciton ${\hat a_s}^\dagger=({\hat
b}^\dagger+{\hat c}^\dagger)/\sqrt{2}$, with energy
$\omega_s=\omega_D-J/2$, and the anti-symmetric exciton ${\hat
a_a}^\dagger=({\hat b}^\dagger-{\hat c}^\dagger)/\sqrt{2}$, with
energy $\omega_a=\omega_D+J/2$. In this basis the interaction
Hamiltonian reads:
\begin{align}
\hat{\mathcal H}_{int} =& - \Big[\Big(\frac g 2 + \tilde g\Big){\hat a}_s^\dagger{\hat a}_s +\Big(\frac g 2 - \tilde g\Big){\hat a}_a^\dagger{\hat a}_a\Big] ({\hat a}+{\hat a}^\dagger)  \label{interaction_processes}\\
\notag &+\frac g 2 ({\hat a} {\hat a}_a {\hat a}_s^\dagger + {\hat a}^\dagger {\hat a}_a^\dagger {\hat a}_s ) + \frac g 2 ({\hat a} {\hat a}_s {\hat a}_a^\dagger + {\hat a}^\dagger {\hat a}_s^\dagger {\hat a}_a ),
\end{align}
\begin{figure}[t!]
\includegraphics[width=0.4\textwidth]{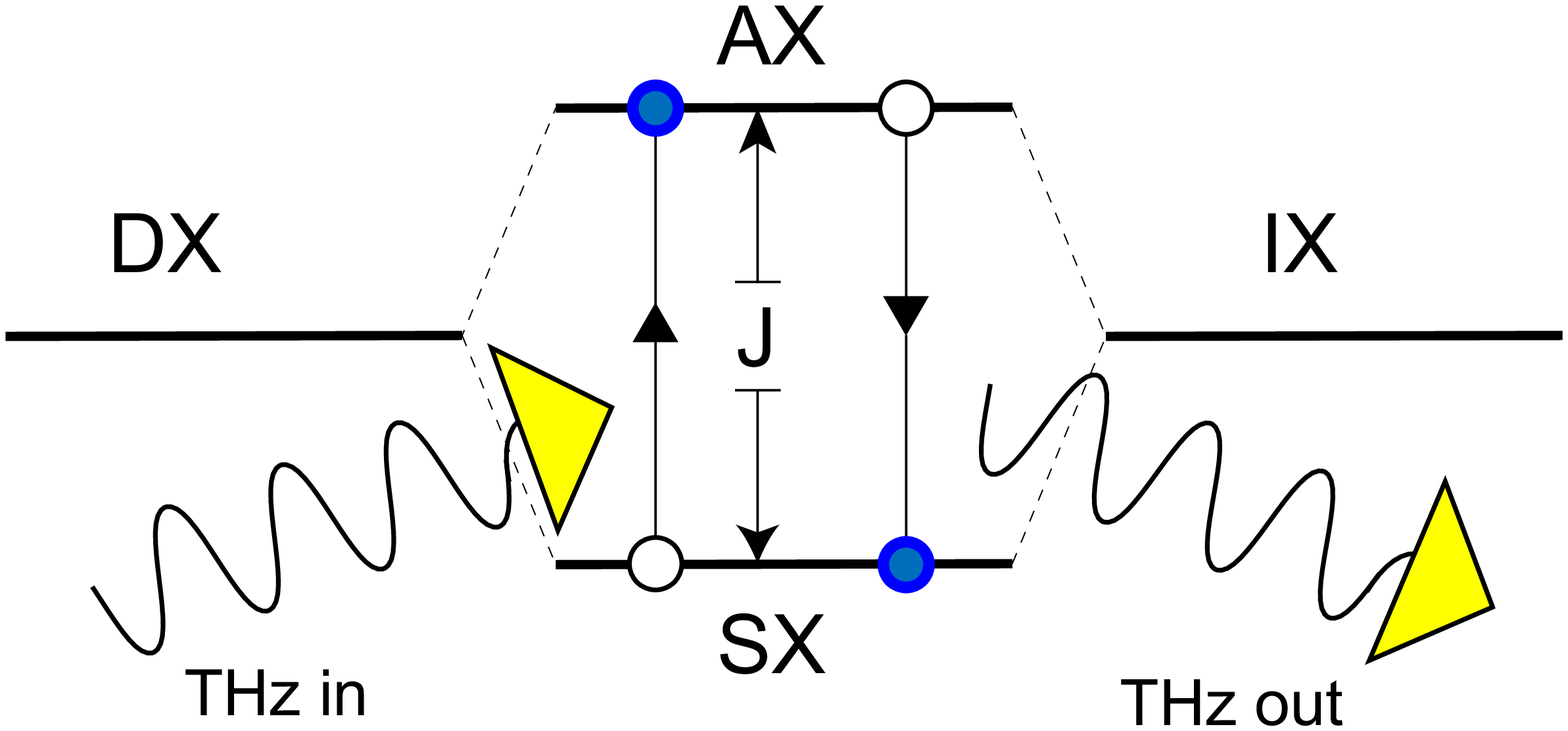}
\caption{(Color online) Energy diagram showing the level repulsion
between the direct (DX) and indirect excitons (IX) at resonance. New
modes, the symmetric (SX) and anti-symmetric exciton (AX) arise,
split by the tunneling rate $J$. The resonant processes in the
interaction Hamiltonian (\ref{interaction_processes}) of THz
absorption and emission are shown.} \label{fig:resonant_processes}
\end{figure}
The first line in Eq. (\ref{interaction_processes}) describes the
interaction of the THz electric field with the static dipole moment
of the excitons. The first term of the second line represents the resonant process
where a THz photon is absorbed and a lower energy symmetric exciton
is excited to the anti-symmetric state, as well as the opposite
process where an anti-symmetric exciton relaxes to the symmetric
state, releasing a THz photon [Fig. \ref{fig:resonant_processes}].
These terms can be expected to give major contribution to THz
emission.  The last term describes the anti-resonant processes which
are usually disregarded using the rotating wave approximation.
However, in our system the transition energies can be in principle
comparable to the coupling constant $g$ and these processes cannot
be in general neglected.

\subsection{Equations of motion and results}

Equations of motion are derived by using the Heisenberg equations
for the annihilation operators, and then calculating the expectation
value defined as $\langle {\hat a_i}\rangle = \text{Tr}\{{\hat \rho}
{\hat a_i}\}$. If we consider the case when occupation numbers are
high, the mean field approximation can be employed, and the following
truncation scheme was used to close the set of dynamic equations,
$\langle {\hat a_i}{\hat a_j} \cdots {\hat a_k} \rangle \approx
\langle{\hat a_i} \rangle \langle{\hat a_j} \rangle \cdots \langle
{\hat a_k}\rangle$ . To remove references to absolute mode energies,
we perform the change of variables ${\hat a_i} \rightarrow
e^{-i\omega_D t} {\hat a_i}$ ($i=$DX,IX). The equations of motion
for mean values of operators read
\begin{align}
\frac{\partial \langle {\hat a} \rangle}{\partial t} =&-i \omega_T \langle{\hat a}\rangle +i\frac {g} \hbar |\langle{\hat c}\rangle|^2+i \frac {2\tilde g} \hbar \text{Re}[\langle \hat b \rangle^* \langle \hat c \rangle] - \frac 1 {2\tau_T} \langle {\hat a} \rangle,
\label{Teqn}\\
\notag
\frac{\partial \langle \hat b \rangle} {\partial t} =&i \frac J 2 \langle \hat c \rangle +i \frac {2\tilde g} \hbar\text{Re}[\langle{\hat a}\rangle]\langle{\hat c}\rangle -i {\tilde P}(t)\\
&- \frac i \hbar (\alpha_1 |\langle \hat b \rangle|^2 + \alpha_2 |\langle \hat c \rangle|^2) \langle \hat b \rangle - \frac 1 {2\tau_{DX}} \langle \hat b \rangle,
\label{DXeqn}\\
\notag
\frac{\partial \langle \hat c \rangle} {\partial t} =&-i\delta_J \langle{\hat c}\rangle + i \frac J 2 \langle \hat b \rangle +i \frac {2} \hbar\text{Re}[\langle{\hat a}\rangle](g\langle{\hat c}\rangle+\tilde g \langle{\hat b}\rangle)  \\
&- \frac i \hbar (\alpha_2|\langle \hat b \rangle|^2 + \alpha_3|\langle \hat c \rangle|^2 )\langle \hat c \rangle- \frac 1 {2\tau_{IX}}\langle \hat c \rangle .
\label{IXeqn}
\end{align}
Lifetimes of the modes have been introduced phenomenologically as
$\tau_{DX} = 1$ ns, $\tau_{IX}=100$ ns and $\tau_{T}= Q/\omega_T$,
where $Q$ is the quality factor of the THz cavity. After the change
of variables the pumping term is written as ${\tilde P}(t) =
e^{i\omega_D t} P(t)/\hbar$. Under CW pumping we write ${\tilde
P}(t) = {\tilde P}_0 e^{-i\Delta_p t}$, where $\Delta_p =
\omega_p-\omega_D$ is the pump detuning from the direct exciton state.

\begin{figure}[t!]
\includegraphics[width=0.48\textwidth]{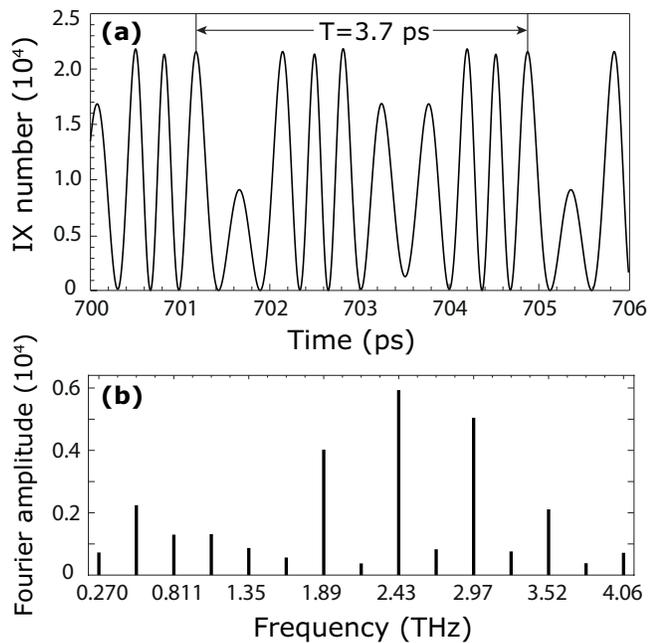}
\caption{(a) Occupation number dynamics of the indirect excitons in
the double QW system under CW pumping. The exciton detuning is
$\hbar \delta_J=1$ meV, and the pump energy is $\hbar\Delta_p=1.5$
meV. Pump strength is linearly turned on to $|{\tilde
P}_0|^2=1.1\cdot 10^{29}$ s$^{-2}$ with a short turning on time,
driving the system into the parametrically instable
regime. The IX occupation number oscillates periodically, but not
harmonically. (b) Spectral characteristics of the time dependence
shown in (a). The oscillations are distributed between many
harmonics, which lowers dramatically the efficiency of THz excitation.}
\label{fig:anharmonic}
\end{figure}
For numerical analysis we used the same set of paremeters as in Ref.
[\onlinecite{Christmann2011}]. The QW$_1$ material is
In$_{0.1}$Ga$_{0.9}$As, the QW$_2$ is grown from
In$_{0.08}$Ga$_{0.92}$As, and the spacer material is GaAs. The well
widths are $d=10$ nm, and the well separation $4$ nm, with the
tunneling rate set to $\hbar J = 6$ meV,\cite{Cristofolini2012}
which allows to estimate the effective electron hole separation as
$L=10$ nm. The direct exciton scattering constant is estimated as
$\alpha_1 = 6 E_b a_B^2/S$,\cite{Tassone} where $a_B=10$ nm is the
direct exciton Bohr radius, and $E_b=8$ meV the binding energy. We
take $S=100$ $\mu$m$^2$ as the system excitation area. The DX-IX
interaction constant is taken from Ref. [\onlinecite{Kristinsson}],
and the indirect exciton scattering constant from the Ref.
[\onlinecite{IXIXint}]. We consider the THz cavity quality factor to be
$Q=100$.\cite{Chasssagneux,Gallant}

In Ref. [\onlinecite{Kristinsson}] it was shown that in the absence
of the THz cavity the equations for coupled direct and indirect
exciton modes display parametric instabilities in the CW pumping
regime. In that work parameters were dressed by the presence of the
resonant optical cavity mode, which decreases dramatically the
direct exciton effective lifetime. A similar behaviour can still be
expected here, and as shown in Fig. \ref{fig:anharmonic}(a), the
exciton numbers are indeed found to oscillate periodically, while
being strongly anharmonic. The corresponding spectrum of the
oscillations is shown in Fig. \ref{fig:anharmonic}(b). One can see that
multiple harmonics have comparable weights, which severely reduces
the coupling of the dipole oscillations to the mode of the THz
cavity. Consequently, the occupation numbers of THz photons remain
extremely small (about $N_{THz} \approx 2$ for the optimal case when
cavity mode is tuned in resonance to transition between symmetric
and antisymmetric states). This makes the anharmonic case very
ineffective for producing single mode THz emission.

\section{DIPOLARITON SYSTEM IN A THZ CAVITY}

In this section we demonstrate that the presence of the optical
cavity tuned close to resonance with excitonic transition can
drastically increase the efficiency of the THz lasing. We consider
the full dipolariton system, consisting of double QWs embedded in a
resonant optical microcavity, with the supplemental THz cavity [see
Fig. \ref{fig:sketch}(b)]. In such a system, the eigenstates are
linear superpositions of the microcavity photon, direct exciton and
indirect exciton, called lower (LP), middle (MP) and upper (UP)
dipolaritons.

\subsection{Hamiltonian and equations of motion}

In the dipolariton system, the QW$_1$ is assumed to be in the strong
coupling regime with the microcavity mode, while the QW$_2$, with
its larger band gap, remains decoupled. The interaction between the
direct and indirect excitons, as well as the interaction of the
indirect exciton with the THz field, remains unchanged. Denoting the
creation operator of the microcavity mode with $\hat a_c^\dagger$,
the Hamiltonian of the new system is
\begin{align}
\notag \hat{\mathcal H} = &\hbar \omega_c {\hat a_c}^\dagger {\hat a_c} + \frac {\hbar \Omega} 2 ({\hat a_c}^\dagger {\hat b} +{\hat b}^\dagger {\hat a_c})+\hat{\mathcal H}_0 +\hat{\mathcal H}_{int}+\hat{\mathcal H}_{nonl} \\
&+P(t){\hat a_c}^\dagger+P(t)^* {\hat a_c}.
\end{align}
The new terms in the first line describe the free propagation of the
microcavity photon with energy $\hbar \omega_c$, and the interaction
between the microcavity mode and the direct exciton described by the
Rabi frequency $\Omega$. The second line describes the coherent
optical pumping, now driving the  optical cavity mode.

Equations of motion are derived in a same way as in the previous
section. The change of variables is performed slightly differently,
as ${\hat a_i} \rightarrow e^{-i\omega_c t} {\hat a_i}$ for modes
$i=$ C, DX and IX. The four coupled equations read
\begin{align}
\frac{\partial \langle {\hat a} \rangle}{\partial t} =&-i \omega_T \langle{\hat a}\rangle +i\frac {g} \hbar |\langle{\hat c}\rangle|^2+i \frac {2\tilde g} \hbar \text{Re}[\langle \hat b \rangle^* \langle \hat c \rangle] - \frac 1 {2\tau_T} \langle {\hat a} \rangle,
\label{Teqn2}\\
\frac{\partial \langle {\hat a_c} \rangle}{\partial t} =& -i \frac{\Omega}{2} \langle \hat b\rangle -\frac{1}{2\tau_C} \langle {\hat a_c} \rangle - i \tilde{P}(t) ,
\label{Ceqn2}\\
\notag
\frac{\partial \langle {\hat b} \rangle} {\partial t} =& i\delta_\Omega \langle \hat b \rangle -i \frac \Omega 2 \langle \hat a_c \rangle +i \frac J 2 \langle \hat c \rangle +i \frac {2\tilde g} \hbar\text{Re}[\langle{\hat a}\rangle]\langle{\hat c}\rangle \\
&- \frac i \hbar (\alpha_1 |\langle \hat b \rangle|^2 + \alpha_2 |\langle \hat c \rangle|^2) \langle \hat b \rangle - \frac 1 {2\tau_{DX}} \langle \hat b \rangle,
\label{DXeqn2}\\
\notag
\frac{\partial \langle \hat c \rangle} {\partial t} =& i(\delta_\Omega - \delta_J)\langle \hat c \rangle + i \frac J 2 \langle \hat b \rangle +i \frac {2} \hbar\text{Re}[\langle{\hat a}\rangle](g\langle{\hat c}\rangle+\tilde g \langle{\hat b}\rangle)  \\
&- \frac i \hbar (\alpha_2|\langle \hat b \rangle|^2 + \alpha_3|\langle \hat c \rangle|^2 )\langle \hat c \rangle- \frac 1 {2\tau_{IX}}\langle \hat c \rangle ,
\label{IXeqn2}
\end{align}
where $\delta_\Omega = \omega_c-\omega_D$. Furthermore we have introduced the decay of the microcavity mode with a lifetime $\tau_c = 5$ ps.

\begin{figure}[t!]
\includegraphics[width=0.48\textwidth]{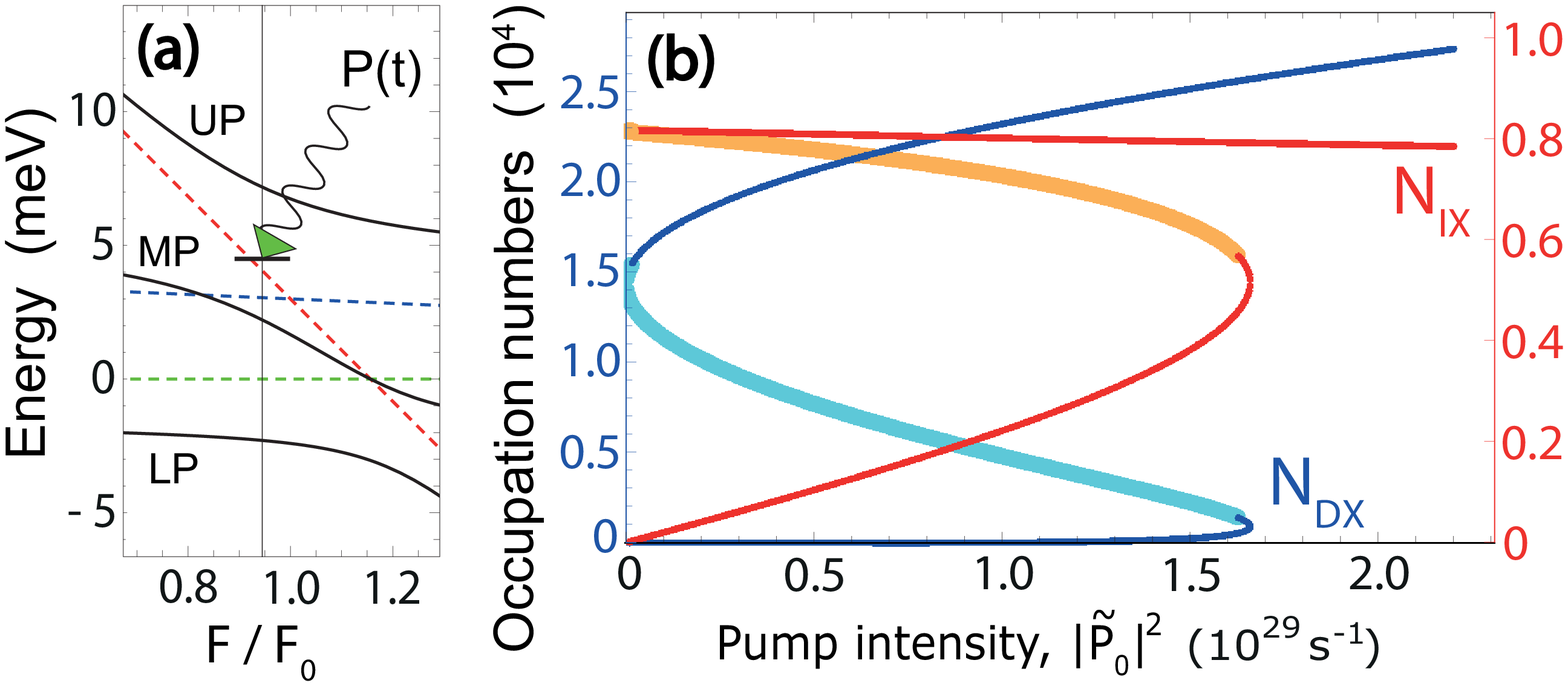}
\caption{(Color online) (a) Energy diagram of the dipolariton system, as a function of the applied field $F$ in units of the resonance field $F_0$. The dashed lines correspond to the bare modes, the photonic (green and flat), the DX (blue) and the IX (red and steep) modes. The diagonalized modes, being the upper (UP), middle (MP) and lower (LP) dipolaritons, are indicated with solid lines. The energy of the pump is identified with a short horizontal line. The vertical line shows the value of the applied field chosen for the calculations. (b) Stability curve under CW pumping for pumping energy $\hbar\Delta_p=4.5$ meV. The bistability arises here due to the blueshift of the MP mode. The blue curve indicates the DX numbers (left axis), and the red curve the IX numbers (right axis). Thin segments indicate stability of the population numbers, while the thick lines indicate parametric instability.}
\label{fig:bistability}
\end{figure}

\subsection{Results}

For the calculations in this section we use the same material
parameters as before, and choose the Rabi splitting as $\hbar \Omega
= 6$ meV.\cite{Cristofolini2012} The tunable parameters are chosen
as  $\hbar \delta_\Omega = -3$ meV and $\hbar \delta_J=1$ meV. The
eigenfrequency of the THz cavity was chosen as $\omega_T/2\pi=1.74$
THz, and quality factor was taken as $Q=100$.

Fig. \ref{fig:bistability}(a) shows the energy diagram of the dipolariton system, with a vertical line indicating the chosen value of the applied field. The energy of the pump lies between the middle and upper dipolariton branches. When the MP mode is populated by the pumping of the system, the corresponding blueshift due to inter-exciton interactions causes bistability, as shown in Fig. \ref{fig:bistability}(b). For these conditions, parametric instability is achieved as indicated in the plot for a wide range of the pump intensities.

\begin{figure}[t!]
\includegraphics[width=0.48\textwidth]{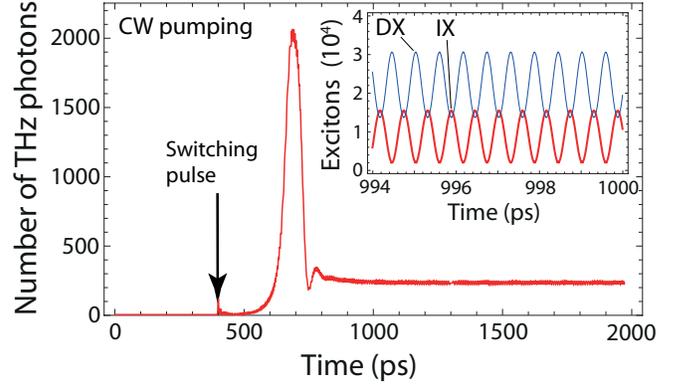}
\caption{(Color online) Plot of the occupation number dynamics of a THz cavity with $Q=100$ coupled to the dipolariton system under CW pumping. The excitonic detuning is $\hbar \delta_J=1$ meV, the cavity mode detuning is $\hbar \delta_{\Omega}=-3$ meV and the relative pump energy is $\hbar \Delta_p=4.5$ meV, corresponding to the energy level configuration in Fig. \ref{fig:bistability}(a). The CW pump is adiabatically turned on to $|{\tilde P}_0|^2=1.1\cdot 10^{29} s^{-2}$. As a result, the system stays on the lower branch of the stability curve depicted in Fig. \ref{fig:bistability}(b). At 400 ps, an additional pulse is applied to the system, switching it to the parametrically instable regime. This provokes oscillations in exciton numbers shown in the inset. For the dipolariton system, these oscillations are harmonic, and result in stable population of the THz mode $N_{THz}\simeq 230$ photons.}
\label{fig:many_THz}
\end{figure}

We solved Eqs. (\ref{Teqn2})-(\ref{IXeqn2}) numerically, considering
the pump which switches on adiabatically to reach its stationary
value. In this situation the system stays on the stable lower branch
of the bistability curve shown at Fig. \ref{fig:bistability}(b). The
exciton numbers are constant in time, the dipole moment is not
oscillating and no THz photons are created, as can be seen in Fig.
\ref{fig:many_THz} for $t<400$ ps. At this moment we apply a short
pulse which switches the system to the parametrically instable state
with oscillations in exciton numbers [see inset to Fig.
\ref{fig:many_THz}] producing high occupancy of the THz mode.

\begin{figure}[b!]
\includegraphics[width=0.48\textwidth]{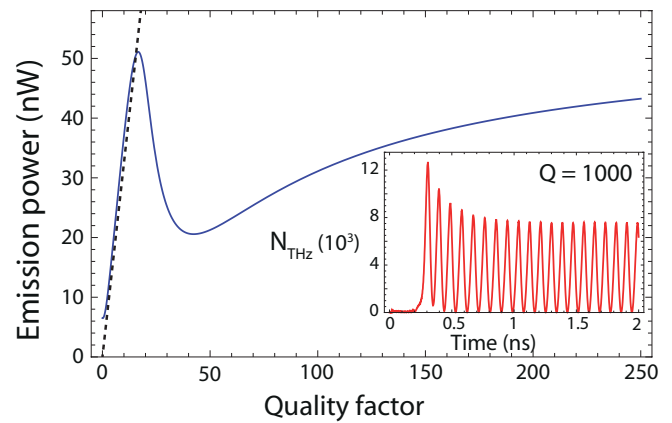}
\caption{(Color online) Plot of the time averaged emission power of THz radiation as a function of the quality factor of the THz cavity. The dashed line shows the low Q Purcell effect prediction in Eq. (\ref{quantumPower}) with the IX oscillation amplitude taken as $N_1=1.1\cdot 10^4$.\cite{Kristinsson} A peak in emission power is found at $Q = 17$, before dropping due to increased feedback of the THz mode on the indirect exciton. The inset shows the number of THz photons as a function of time for $Q=1000$. For such a high quality factor the number of THz photons is considerably large. The resulting feedback on the excitonic modes is so strong it does not allow for a steady state solution. Rather, the number of THz photons oscillates, with a period in the range of 200 ps, mimicking the output of a Q-switched THz laser.}
\label{fig:Q-factor}
\end{figure}

After a transitory period of about $\sim$400 ps during which a
strong outburst of THz radiation occurs, oscillations  become
harmonic enough for the THz occupancy to stabilize at the value
$N_{THz}\simeq 230$, which is about two orders of magnitude greater
then those obtained in the situation where the optical cavity is
absent. This drastic increase is connected with the fact that the
presence of the optical cavity makes the oscillations of the exciton
occupancies highly harmonic [see Fig. \ref{fig:many_THz}, inset],
which is favorable for monomode emission. This produces constant THz
lasing from the system due to the escape of the THz photons out of
the cavity. For the chosen parameters the power of the laser can be
estimated as $I_0= N_{THz} \hbar \omega_T /\tau_T \simeq 30\text{
nW}$. The dependence of the emission power on the quality factor of the
THz cavity is demonstrated in Fig. \ref{fig:Q-factor}.

For small $Q$ one can assume that occupancy of the THz mode is
relatively small, and its presence only slightly modifies the
oscillations between direct and indirect excitons. The equation for
the THz mode can be then decoupled from other equations and be
considered as linear differential equation with external pumping
term. If we neglect the effect of the smaller $\tilde g$ term we
have
\begin{align}
\frac{\partial \langle {\hat a} \rangle}{\partial t} =&-\Big(i \omega_T+ \frac 1 {2\tau_T}\Big)\langle{\hat a}\rangle +i\frac {g} \hbar N_{IX}(t),
\end{align}
where the occupancy of the indirect exciton mode can be to very high precision approximated by a harmonic function:
\begin{equation}
N_{IX}(t)=N_0 + N_1 (e^{i\omega t}+e^{-i\omega t})/4, \label{IXnumber}
\end{equation}
where $N_0$ is the average occupancy, and $N_1$ is the
peak-to-valley amplitude of the IX oscillations. The third term is
the only resonantly driving term, and we neglect the other two,
assuming that the frequency of the excitonic oscillations is in
resonance with the eigenfrequency of the THz cavity,
$\omega=\omega_{T}$. The stationary solution of the resulting
equation gives for the occupancy of the THz mode:
\begin{align}
N_{THz}=|\langle \hat a\rangle|^2= \frac{g^2 N_1^2 \tau_T^2}{2\hbar^2},
\end{align}
which corresponds to the emission power of
\begin{align}
I_0^q = \frac{N_{THz}} {\tau_T} \hbar \omega_T = \frac{g^2}{4\hbar}N_1^2 Q.\label{quantumPower}
\end{align}
An important attribute of Eq. (\ref{quantumPower}) is that the
emitted power is proportional to the square of the indirect exciton
number. This corresponds to the \textit{superradiance}
effect,\cite{Dicke,Bohnet} for which the coherence of the quantum
mechanical oscillators causes emission to increase superlinearly
with the number of oscillators. This phenomenon allows reaching high
emission power at the engineering stage by up-scaling of the
device, making it competitive with other schemes of THz generation.

In Refs. [\onlinecite{Kyriienko2013}] and [\onlinecite{Kristinsson}] the emission power for the dipolariton system was estimated classically as
\begin{align}
I_0^c = \frac{N_1^2 d_0^2 \omega_T^4}{48\pi \varepsilon c^3}, \label{classicalPower}
\end{align}
By dividing the quantum mechanical estimate with the classical estimate, one immediately sees that the presence of THz cavity increases the intensity of the emission by the Purcell factor of the system:
\begin{align}
F_P = \frac{I_0^q}{I_0^c} = \frac 3 {4\pi^2} \frac{\lambda_T^3}{V} Q,
\end{align}
where $\lambda_{T}$ is a wave length of THz radiation.
\begin{figure}[t!]
\includegraphics[width=0.4\textwidth]{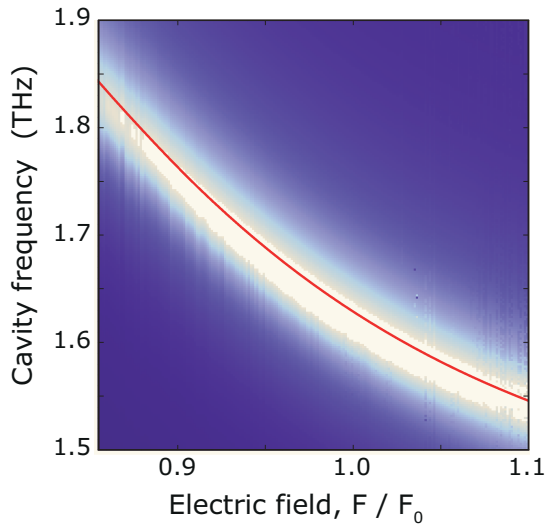}
\caption{(Color online) Density plot of the emission power as a function of the applied field and cavity eigenfrequency. The optical cavity detuning is $\hbar\delta_\Omega=-1$ meV, the pump energy is $\hbar\Delta_p = 4.5$ meV, and the pumping strength is $|\tilde P _0|^2=1.1\cdot 10^{29}$ s$^{-2}$. The quality factor of the THz cavity is $Q=17$, corresponding to the optimal emission power in Fig. \ref{fig:Q-factor}. Lighter colors signify higher output power, with the maximum output power following closely the indirect exciton oscillation frequency (red solid line).}
\label{fig:resonance_plot}
\end{figure}

The linear dependence of the emission power on quality factor given
by Eq. (\ref{quantumPower}) is shown by a dotted line in Fig.
\ref{fig:Q-factor}. It gives a good approximation for the exact
curve for quality factors up to $Q\simeq 17$.
Past this point, the emission power experiences a drop due to the increased feedback of the THz mode on excitonic oscillations and then starts to rise again. Thus, for the chosen parameters the relatively low Q-factor of 17 is optimal for the THz lasing based on the dipolariton system.

For very large quality factors, $Q\gtrsim 800$, there appear
oscillations in the THz photon occupancy (see the inset to Fig.
\ref{fig:Q-factor}). The high number of THz photons that are quickly
excited results in a strong enough feedback to disrupt the resonant
oscillations of indirect excitons. As THz photons generation is
consequently suppressed, their occupancy decays, and the feedback on
the IX mode disappears. At this point exciton oscillations
restabilize, again exciting a large number of THz photons. As a
steady state THz occupancy does not develop, it results in the
periodical modulation of emission power. This mimics the
functionality of a Q-switched laser, with higher peak power than
that of the lower Q-factor CW emission.

An important question to address is the tunability of the emission.
The period of the oscillations of the indirect exciton numbers can
be changed by the applied field,\cite{Kyriienko2013,Kristinsson}
which alters the energy level structure of the system [see Fig.
\ref{fig:bistability}(a)]. Tuning the system in this manner will
move the system out of resonance with the THz cavity, lowering the
emission power. However, the THz cavity eigenfrequency can be
imagined to be alterable, for instance by applying stress to deform
the cavity. By simultaneously changing the applied field and the
eigenfrequency of the THz cavity, the frequency of emission can be
tuned, while staying in resonance.

A surface plot of the emission power is presented in Fig.
\ref{fig:resonance_plot} as function of the applied field and the
THz cavity frequency. The highest emission power follows closely the
red solid line, which indicates the frequency of IX oscillations,
following the same electric field dependence as previously observed
in Ref. [\onlinecite{Kristinsson}]. We find that maximum output
power drops in the low electric field limit [see Fig.
\ref{fig:resonance_plot}], caused by the shift of multistability
region to the lower pumping strengths, with pump intensity being
kept fixed. Consequently, higher values of lasing frequencies can be
achieved by tuning the pumping strength. Going beyond the electric
field strength shown in Fig. \ref{fig:resonance_plot} the output
power greatly diminishes, since the system enters the parameter
regime where indirect exciton number oscillations become anharmonic.

\section{CONCLUSIONS}
We have developed a microscopic theory of the terahertz lasing from
the dipolariton system. We have shown that in the case of simple
double quantum wells embedded in a THz cavity, the tunneling between
direct and indirect exciton modes does not lead to stable THz
lasing. The presence of an optical cavity strongly coupled with
the direct exciton can improve the situation, and stable THz emission
becomes possible. The output power of emission was analysed as
function of THz cavity parameters. In particular, we showed that the
THz emission power has a peculiar dependence on THz cavity Q-factor,
showing an optimal value of about $17$ for powerful CW emission.
The effect of THz emission superradiance was discussed as a way to
achieve high output power. Additionally, we revealed the Q-switched
behavior of THz lasing for a high finesse Thz cavity driven by
dynamic feedback effects.

\begin{acknowledgements}
We thank Timothy C. H. Liew for useful discussions on the subject.
This work has been supported by FP7 IRSES projects ``POLATER'' and
``POLAPHEN'', and Tier1 project ``Novel polaritonic devices''. O. K.
acknowledges the support from Eimskip Fund.
\end{acknowledgements}

\appendix

\section{Overlap dipole matrix element}

In this section we calculate the dipole matrix element
$d_{di}=-e\langle DX|\hat{z}|IX\rangle$. This can be done by
calculating the exchange integral between the ground states of two
finite potential quantum wells. The time-independent Schr\"odinger
equations for each separate well read as
\begin{align}
E\psi(z) = -\frac{\hbar ^2}{2m^*} \psi''(z)+V(z)\psi(z), \label{Schr}
\end{align}
where
\begin{align}
V(z)=V_0\Big(\theta(-z-d/2)+\theta(z-d/2)\Big),
\end{align}
$V_0$ is the QW depth, and $d$ is the well width. The ground state solution to Eq. (\ref{Schr}) reads
\begin{align}
\psi(z)=\begin{cases}
Ae^{\alpha z} &\text{ if }z<-L/2 \\
B\cos(kz) &\text{ if }|z|<L/2 \\
Ae^{-\alpha z} &\text{ if }z>L/2
\end{cases},
\end{align}
where
\begin{align*}
\alpha &= \sqrt{2m^*(V_0 -E)}/\hbar, \\
k &= \sqrt{2m^*E}/\hbar,
\end{align*}
are parameters found from the lowest $k$ solution to the equation
\begin{align}
\sqrt{2m^*V_0-\hbar^2 k^2 } = \hbar^2 k^2\tan(k d/2).
\end{align}
Normalization of the states gives coefficients
\begin{align}
B&=\Big(\frac{kd + \text{sin}(kd)}{2k}+\frac{\text{cos}(kd/2)^2}{\alpha}\Big)^{-1/2},\\
A&=\text{cos}(kd/2)e^{\alpha d/2} B.
\end{align}
In addition to the parameters in the main text, the valence band offset between the GaAs bulk and the In$_{0.1}$Ga$_{0.9}$As/In$_{0.08}$Ga$_{0.92}$As QW is $V_0 = 55$ meV.\cite{Komsa} Finally, the exchange integral can be straightforwardly calculated as
\begin{align}
\notag d_{di} &= \langle DX |(-e\hat z)|IX\rangle \\
&=-e\int \psi(z) z \psi(z-D) dz \approx -0.17 eL,
\end{align}
where $D=14$ nm is the distance between the centers of the two QWs.

\end{document}